\documentclass[twocolumn,prd,showpacs]{revtex4}
\usepackage{graphicx}
\usepackage{dcolumn}
\usepackage{bm}

\newcommand {\bd}{\begin{displaymath}}
\newcommand {\ed}{\end{displaymath}}
\newcommand {\eq}{\begin{equation}}
\newcommand {\beq}{\begin{equation}}
\newcommand {\eeq}{\end{equation}}
\newcommand {\beqa}{\begin{eqnarray}}
\newcommand {\eeqa}{\end{eqnarray}}
\newcommand {\n}{\nonumber \\}
\newcommand {\tr}{{\rm tr\,}}

\newcommand {\RR}{\mbox{\scriptsize R}}
\newcommand {\II}{\mbox{\scriptsize I}}

\newcommand {\trs}{\mbox{\scriptsize tr}}

\newcommand {\ee}{\mbox{e}}

\newcommand {\dd}{\mbox{d}}

\newcommand {\del}{\partial}

\newcommand {\defeq}{\stackrel{\rm def}{=}}

\newcommand {\vev} [1] {\langle #1 \rangle}
\newcommand{\id}{{1\!\!1}} 

\begin{document}

\preprint{NBI--HE--03--XX}

\title{Non-Commutativity of the Zero Chemical Potential Limit\\
and the Thermodynamic Limit in 
Finite Density Systems}
\author{J.\ Ambj\o rn}
\email{ambjorn@nbi.dk}
\affiliation{%
The Niels Bohr Institute, Blegdamsvej 17, 2100 Copenhagen, Denmark}
\affiliation{%
Institute for Theoretical Physics, Utrecht University,
Leuvenlaan 4, 3584 CE Utrecht, The Netherlands}
\author{K.N.\ Anagnostopoulos}
\email{konstant@physics.uoc.gr}
\affiliation{%
Department of Physics, University of Crete,
P.O. Box 2208, GR-71003 Heraklion, Greece
}
\affiliation{%
Physics Department, National Technical University, Zografou
Campus, GR-15780 Athens, Greece}
\author{J.\ Nishimura}
\email{jnishi@post.kek.jp}
\affiliation{%
High Energy Accelerator Research Organization (KEK), 1-1 Oho, Tsukuba
306-0801, Japan
}%
\author{J.J.M.\ Verbaarschot}
\email{verbaarschot@tonic.physics.sunysb.edu}
\affiliation{%
Department of Physics and Astronomy, SUNY, Stony Brook, NY 11794, USA}

\date{\today}

\begin{abstract}
Monte Carlo simulations of finite density systems are often
plagued by the complex action problem.
We point out that there exists certain
non-commutativity in the zero chemical potential limit and the
thermodynamic limit when one tries to study such systems by
reweighting techniques.
This is demonstrated by explicit calculations in a Random Matrix Theory, 
which is thought to be a simple qualitative model for finite density QCD.
The factorization method allows us to understand
how the non-commutativity, which appears at the intermediate steps,
cancels in the end results for physical observables.

\end{abstract}

\maketitle

\section{\label{sec:intro}Introduction}
QCD at finite baryon density and/or finite temperature
has attracted much attention 
due to its relevance to the physics
of the early universe, heavy ion collisions and neutron stars.
It is of particular importance 
to explore the phase diagram
in the $\mu$ (chemical potential), $T$ (temperature) plane,
where interesting phases such as a superconducting phase
have been conjectured to appear. 
Monte Carlo simulations, which enable first-principle studies
at $\mu = 0$, are hindered by
the fact that the fermion determinant becomes complex for
$\mu \neq 0$. The standard reweighting method uses the absolute value
for generating configurations, and takes account of 
the phase in measuring observables.
Due to cancellations caused by the oscillating phase, however,
the required number of configurations grows
exponentially with the system size.
Furthermore the method suffers from the so-called overlap problem,
which is caused by the mismatch of the region in the configuration space
which contributes to the ensemble average and the region which one mostly
samples in the phase-quenched model.

One of the recent developments in this direction is
that the reweighting techniques have proven to
be of use in exploring the phase diagram at small $\mu$
and large $T$, where the fluctuation of the phase is still under control.
Substantial improvements are achieved by various tricks 
such as the multi-parameter reweighting \cite{Fodor:2001au,Fodor:2001pe},
Taylor expansion approach \cite{Allton:2002zi}
and the imaginary $\mu$ approach \cite{deForcrand:2002ci,D'Elia:2002gd}.
In particular the first approach was able to locate the
critical end point in the $\mu$-$T$ plane \cite{Fodor:2001pe}.
(See also Ref.\ \cite{Ejiri:2004yw}.)

In Ref.\ \cite{sign} two of the authors (K.N.A.\ and J.N.)
proposed a new method, the factorization method,
for simulating systems with complex actions.
The method utilizes the fact that
the distribution of observables factorizes into the corresponding 
distribution for the phase-quenched model and the weight factor
representing the effect of the phase.
Each factor can be obtained by constrained Monte Carlo simulations,
which eliminate the overlap problem completely.
The knowledge of the weight factor allows us to understand the
effect of the phase intuitively.
The method is quite general,
and in particular it is expected
to be useful in going beyond the small $\mu$ regime in finite density QCD.
The method proposed for simulating $\theta$-vacuum like systems
\cite{Azcoiti:2002vk} may be viewed as a particular case of the
factorization method.
In Ref.\ \cite{Muroya:2003qs} it was pointed out that
the factorization method belongs to the class of methods known as
`the density of states method'.

In Ref.\ \cite{rmt} we have tested the method in
a Random Matrix Theory (RMT), which is thought of
as a schematic model for QCD at finite baryon density \cite{Stephanov:1996ki}.
RMT was
originally introduced to describe the spectrum of the Dirac operator
and has been extensively studied in the 
literature \cite{Verbaarschot:2000dy}.
The model we have studied exhibits
a first order phase transition at some value of the ``chemical
potential'' and 
can be solved analytically even for finite size matrices \cite{Halasz:1997he}.
One hopes to capture the essential
properties of real QCD, while at the same time having a testing ground
\cite{Halasz:1999gc} for methods to be applied to real QCD.
Indeed the factorization method is quite successful in RMT \cite{rmt},
where exact results are reproduced with great accuracy,
and one understands clearly
how the phase of the determinant
induces the first order phase transition.

In this paper we investigate this model further in order to address the
question of the non-commutativity in the $\mu \rightarrow 0$
limit and the thermodynamic limit.
The factorization method allows us to 
understand how the non-commutativity appears at the intermediate steps of
reweighting techniques, and how it cancels 
in the end results for physical observables.
Preliminary results have already been presented in
conference proceedings, Refs.\ \cite{ncproc,Nishimura:2003ri}.


\section{RMT for Finite Density QCD and the Monte Carlo Methods}
\label{sec:model}

We consider the RMT defined by the partition function
\beq
Z = \int \dd W \ee ^{- N \, \trs (W^\dag W)} \, \det D  \ ,
\label{rmtdef}
\eeq
where $W$ is a $N \times N$ complex matrix, and 
$D$ is a $2N \times 2N$ matrix given by
\beq
D = 
\left( 
\begin{array}{cc}
m & i W +  \mu  \\
i W^\dag  + \mu & m
\end{array}
\right) \ .
\label{defD}
\eeq
The above model can be thought of as a schematic model of finite
density QCD with one flavor, where the parameters $\mu$ and $m$
correspond to the chemical potential and the quark mass respectively.
The size of the matrix $W$ corresponds to the number
of low lying modes of the Dirac operator and if the density of these
modes is taken to be unity, $N$ can be interpreted as the volume of
space-time. As physical observables, one may consider the 
``chiral condensate'' and the ``quark number density'' defined by 
\beqa
\Sigma &=&  \frac{1}{2N} \, \tr \, (D^{-1})  \ , \\
\nu   & = &   \frac{1}{2N} \, \tr \, ( \gamma_4 D^{-1} ) \ ,
\mbox{~~~~~~~} \gamma_4 = \left( 
\begin{array}{cc}
0 & \id  \\
\id  & 0
\end{array}
\right) \ .
\eeqa
In what follows we consider the massless case ($m=0$) and
focus on the ``quark number density''.

\begin{figure}[htbp]
  \begin{center}
    \includegraphics[width=8cm]{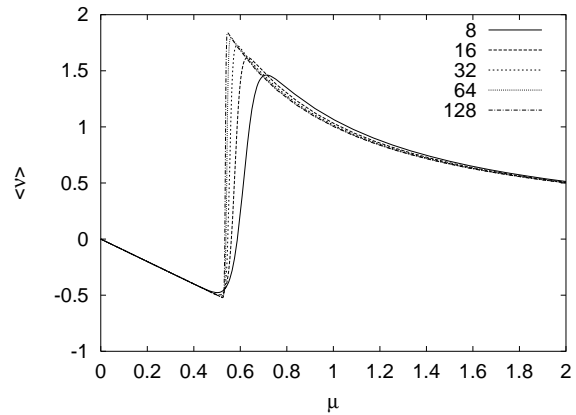}
    \caption{The exact result (\protect\ref{finiteN}) 
for the `quark number density' $\langle \nu \rangle$ is 
plotted as a function of the `chemical potential' $\mu$
for $N=8,16,32,64,128$.
In the $N \rightarrow \infty$ limit, the function develops
a discontinuity at $\mu = \mu_{\rm c}=0.527\cdots$.
}
    \label{fig:nuser}
  \end{center}
\end{figure}

The model was first solved in the large-$N$ limit
\cite{Stephanov:1996ki}, but an analytic solution has been obtained in  
\cite{Halasz:1997he} even for finite $N$. The partition function can
be expressed as 
\beq
Z(\mu) = \pi \ee^{\kappa} N ^{-(N+1)} \, N !
\left[ 1+ \frac{(-1)^{N+1}}{N !} \gamma (N+1 ,\kappa)
\right] \ ,
\label{partitionfn}
\eeq
where $\kappa = - N \mu^2$ and $\gamma(n,x)$ is the incomplete 
$\gamma$-function defined by
\beq
\gamma(n,x)=\int_0 ^x \ee^{-t} \, t^{n-1} \, dt \ .
\eeq
From this one obtains the vacuum expectation value 
(VEV) of the quark number density as
\beqa
\langle \nu \rangle
&=& \frac{1}{2N} \frac{\del}{\del \mu} \ln Z(\mu) \\
&=& - \mu \left[ 1 + \frac{\kappa ^N \ee^{-\kappa}}
{ (-1)^{N+1}N\! + \gamma(N+1 , \kappa)   }
\right] \ .
\label{finiteN}
\eeqa
In Fig.\ \ref{fig:nuser} we plot $\langle \nu \rangle$ as a function
of the chemical potential $\mu$ for $N=8,16,32,64,128$.  The large-$N$
limit of this formula 
is easily found
by applying the saddle-point
method to the incomplete $\gamma$-function. We obtain
\beq
\lim_{N\rightarrow \infty}
 \langle \nu \rangle =
\left\{ \begin{array}{ll}
- \mu  & \mbox{for~}\mu < \mu _{\rm c}  \\
1/ \mu  & \mbox{for~}\mu > \mu _{\rm c} \ ,
\end{array} 
\right. 
\label{nq_rmt}
\eeq
where $\mu_{\rm c}$ is the solution to the equation $1 + \mu^2 + \ln
(\mu^2) = 0$, and its numerical value is given by $\mu_{\rm c}
=0.527\cdots$.  We find that the quark number density $\langle \nu
\rangle$ has a discontinuity at $\mu = \mu_{\rm c}$.  Thus the
schematic model reproduces qualitatively the first order phase
transition expected to occur in `real' QCD at nonzero baryon density.

The model (\ref{rmtdef})
cannot be simulated directly because the fermion determinant
has a non-zero phase $\Gamma$ defined by
\beq
\det D =  \ee ^ {i \Gamma}\,  | \det D | \ ,
\eeq
when $\mu \neq 0$.
In order to reveal the importance of the phase
let us consider the {\em phase-quenched} model
\beqa
\label{part_absdef}
Z_0 &=& \int \dd W \, \ee ^{- N \, \trs (W^\dag W)} \, | \det D |  \n
&=& \int \dd W \, \ee ^{-S_0} \ , \\
S_0 &=& N \, \tr (W ^\dag W) - \ln | \det D |  \ ,
\label{absdef}
\eeqa
and the corresponding VEVs, which we denote as $\vev{\ldots}_0$.
The large-$N$ limit of $\vev{\nu}_0$ can be obtained analytically
as \cite{Stephanov:1996ki}
\beq
\lim_{N\rightarrow \infty}
\langle \nu \rangle_0 = 
\left\{ \begin{array}{ll}
\mu  & \mbox{for~}\mu < 1  \\
1/ \mu   & \mbox{for~}\mu > 1 \ .
\end{array} 
\right. 
\label{nq_abs}
\eeq 
Comparing this with the result (\ref{nq_rmt}) for the full model,
we see that the effect of $\Gamma$ is dramatic for
$\mu < 1$.

The reweighting method uses the identity
\beq
\left\langle \nu \right\rangle
= \frac{\left\langle \nu \, \ee ^{i \Gamma }
\right\rangle _{0}}
{\left\langle \ee ^{i \Gamma }
\right\rangle _{0}} 
\label{VEV}
\eeq
to calculate $\left\langle \nu \right\rangle$ for the full model 
by Monte Carlo simulation of the phase-quenched model.
Due to the symmetry of the phase-quenched model
under $W \mapsto - W$, where the fermion determinant $\det D$ 
as well as the observable $\nu$ becomes complex conjugate, we obtain
\beqa
\label{sum_nu}
 \langle \nu \rangle &=& \langle \nu_{\rm R} \rangle
+ i \, \langle \nu_{\rm I} \rangle  \ , \\
  \langle \nu_{\rm R} \rangle &=&
\frac{\left\langle \nu_{\rm R} \cos \Gamma 
\right\rangle _{0}}
{\left\langle \cos \Gamma 
\right\rangle _{0}}   \ ,  
\label{reweightR}
\\
\langle \nu_{\rm I} \rangle &=&
i \, \frac{\left\langle \nu_{\rm I} \sin \Gamma 
\right\rangle _{0}}
{\left\langle \cos \Gamma 
\right\rangle _{0}}  \ ,
\label{reweight}
\eeqa
where $\nu_{\rm R}$ and $\nu_{\rm I}$ denote the real part
and the imaginary part of $\nu$, respectively. The same symmetry
implies that 
\beq
\langle \nu _{\rm R} \rangle _0 =\langle \nu \rangle _0 
~~~;~~~ \langle \nu_{\rm I}\rangle _0 = 0 \ .
\label{RI0}
\eeq
For the unquenched model, on the other hand, one obtains \cite{rmt}
\beq
\langle \nu_{\rm R} \rangle = 0 
 ~~~;~~~
\langle \nu_{\rm I} \rangle = i \, \mu
\eeq
in the large $N$ limit for $\mu < \mu_{\rm c}$, 
which is in sharp contrast to the phase-quenched result (\ref{RI0}).

The factorization method in the present case amounts to calculating the
VEVs on the r.h.s.\ of 
(\ref{reweightR}) and (\ref{reweight}) by the formulae
\beqa
\label{fact1}
\left\langle \nu_{\rm R} \cos \Gamma 
\right\rangle _{0} &=& 
 \int _{-\infty} ^{\infty}
 \dd x \, x \, \rho_{\rm R}^{(0)} (x) \, w_{\rm R}(x) \ ,\\
\label{fact2}
\left\langle \cos \Gamma 
\right\rangle _{0}
&=&
\int _{-\infty} ^{\infty} \dd x 
\, \rho^{(0)} _{\rm R} (x) \,  w _{\rm R} (x)   \ , \\
\label{fact3}
\left\langle \nu_{\rm I} \sin \Gamma 
\right\rangle _{0}
&=& 
2 \int  _{0} ^{\infty}
\dd x \, x \, \rho ^{(0)}_{\rm I} (x) \, w_{\rm I} (x) \ ,
\eeqa
where the functions $\rho^{(0)}_i (x)$ ($i={\rm R,I}$)
represent the distribution of $\nu_i $ ($i={\rm R,I}$)
in the phase-quenched model
\beq
\rho^{(0)}_i (x) \defeq \langle \delta (x - \nu_i )\rangle_0
 \ .
\label{constraint_model}
\eeq
The weight factors $w_i(x)$ in Eqs.\
(\ref{fact1})--(\ref{fact3}) are defined by
\beqa
\label{wRcos}
w_{\rm R}  (x) &\defeq& \langle  \cos \Gamma \rangle_{\RR , x }  \ ,\\
w_{\rm I}(x) &\defeq& \langle \sin \Gamma\rangle_{\II , x}  \ ,
\label{wIsin}
\eeqa
where the VEV $\vev{\ldots}_{i,x}$ is taken with respect to yet another
partition function
\beq
Z_i(x) = \int \dd W \, \ee^{-S_0} \, \delta(x-\nu_i)  \ .
\label{cnstr_part}
\eeq

The problem then reduces to the calculation
of the four functions
$\rho^{(0)}_i (x)$ and $w_i(x)$ ($i={\rm R,I}$).
This may be done in various ways. In Ref.\ \cite{rmt} as well as
in the present work we have simulated
\beq
Z_{i,V}
 =  \int \dd W
 ~ \ee^{-S_0 }~ \ee ^{- V(\nu_i) }  \ ,
\label{part_pot}
\eeq
where the $\delta$-function in (\ref{cnstr_part}) is replaced
by a sharply peaked potential $V(x)$, which is taken to be Gaussian 
\beq
V(x) = \frac{1}{2} \gamma (x - \xi )^2
\label{pot}
\eeq
with $\gamma$ and $\xi$ being real parameters. By choosing $\gamma$ large
enough, the results become insensitive to its value (we used $\gamma =
1000.0$). The functions $\rho_i^{(0)} (x)$ can be
obtained from the same simulation. We refer the reader to Ref.\ \cite{rmt}
for more details.

The complex action problem occurs in the reweighting method
because the trigonometric functions in (\ref{reweightR}) and (\ref{reweight})
flip their signs violently when one moves around the configuration space.
The same problem occurs in the factorization method when one calculates
the weight factors (\ref{wRcos}), (\ref{wIsin}). However, by
simulating the constrained system (\ref{part_pot}),
one forces the simulation to sample the important region of the configuration
space, which is rarely visited by the simulation of 
the phase-quenched model (\ref{part_absdef}).
Thus the factorization method removes the overlap problem completely.
Once the weight factors are obtained roughly, one can
make the sampling more efficient by the use of multi-canonical simulations.
This is not yet done, however.
The knowledge of the weight factors is also useful in understanding
the effect of the phase intuitively \cite{sign,rmt},
and it plays a crucial role in the present work.

\section{Non-Commutativity of the $\mu\to 0$ and the Thermodynamic Limit}
\label{sec:results}

In this Section we discuss
the non-commutativity of 
the two limits $\mu \rightarrow 0$ and $N \rightarrow\infty$
in the RMT.
Such non-commutativity can be readily seen from
the partition function (\ref{partitionfn}).
Below the critical point, the large $N$ behavior
is given by $Z(\mu) \sim {\rm const.} \ee^{- N\mu^2}$.
Omitting the $\mu$-independent factor,
the partition function approaches unity if one takes the 
$\mu \rightarrow 0$ limit first,
whereas it vanishes if one takes the large $N$ limit first.
More generally, one obtains $\ee^{-C}$, if one takes
the two limits simultaneously with fixed $N \mu^2 \equiv C$.
This non-commutativity is caused by the phase of the determinant.
The phase vanishes at $\mu = 0$ for
finite $N$, and one obtains a nonzero result for the partition
function in the large $N$ limit with appropriate normalization.  
If one takes the large $N$ limit first for small but finite $\mu$, however,
the oscillation of the phase becomes violent,
and as a result one obtains $Z=0$ with the same normalization.

We note, however, that the free energy
defined by 
\beq
 f(\mu) = - \lim_{N\rightarrow \infty}
\left\{ \frac{1}{N} \ln Z(\mu , N) - {\rm const.} \right\}
\eeq
after subtracting the $\mu$-independent constant
is given by $f(\mu) = \mu^2$ at $\mu < \mu_{\rm c}$,
which is continuous at $\mu = 0$
\footnote{We take this opportunity to correct 
the proceedings \cite{Nishimura:2003ri},
where it was wrongly stated that the free energy {\em has} a discontinuity.}.
Furthermore we find 
from (\ref{finiteN}) or Fig.\ \ref{fig:nuser}
that the quark number density does not
have such non-commutativity, either.
Thus the non-commutativity does not seem to appear in physical 
quantities, but it appears at the intermediate steps
of the reweighting method as we will see in what follows.

\begin{figure}[htbp]
  \begin{center}
    \includegraphics[width=8cm]{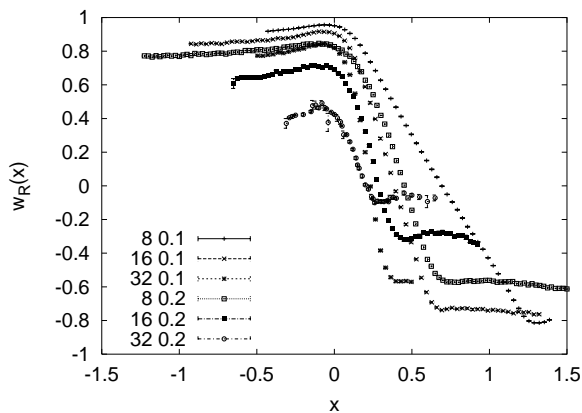}
    \caption{
The weight factor $w_{\rm R}(x)$ is plotted
for $\mu = 0.1$ and 0.2 at $N= 8,16,32$.
}
    \label{fig:rhoR}
  \end{center}
\end{figure}

Let us first consider the weight factor $w_{\rm R}(x)$,
which has a non-commutativity similar to the partition function.
Since the phase $\Gamma$ disappears identically for $\mu = 0$,
one obtains $w_{\rm R}(x)\equiv 1 $ at $\mu = 0$ for any $N$.
On the other hand, one obtains $w_{\rm R}(x) \equiv 0$ 
in the large $N$ limit for any $\mu \neq 0$,
since the phase oscillates violently. 
In Fig.\ \ref{fig:rhoR} we plot $w_{\rm R}(x)$ for $\mu = 0.1$
and $\mu = 0.2$ at $N=8,16,32$.
The weight factor $w_{\rm R}(x)$ crosses zero, 
and the crossing point moves to infinity as $\mu \rightarrow 0$.
Thus the convergence of $w_{\rm R}(x)$ to $w_{\rm R}(x)\equiv 1 $
in the $\mu \rightarrow 0$ limit is {\em not} uniform.

\begin{figure}[htbp]
  \begin{center}
    \includegraphics[width=8cm]{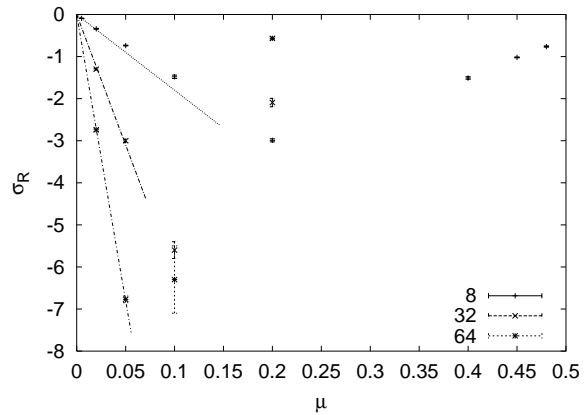}
    \caption{
The slope $\sigma_{\rm R}(\mu,N)$
of the weight factor $w_{\rm R}(x)$ in the linear regime
is plotted against $\mu$ for $N= 8,32,64$. 
For sufficiently small $\mu$
it fits well to $\sigma_{\rm R}(\mu,N) \sim - \alpha_{\rm R} (N) \, \mu$.
}
    \label{fig:slopeR}
  \end{center}
\end{figure}

As $\mu$ approaches zero for fixed $N$, a linear regime 
develops in the region where $w_{\rm R}(x)$ crosses zero.
We extract the slope $\sigma_{\rm R} (\mu , N)$
in this regime and plot it against $\mu$ in Fig.\ \ref{fig:slopeR}.
At small $\mu$ we find that the slope can be fitted nicely by
\beq
\sigma_{\rm R} (\mu , N) \sim - \alpha_{\rm R} (N)  \, \mu  \ ,
\label{sigmaR_asym}
\eeq
where the coefficient $\alpha_{\rm R} (N)$ grows linearly with $N$.
Therefore the asymptotic behavior of the 
weight factor $w_{\rm R} (x)$ depends
on how the two limits $\mu \rightarrow 0 $, $N \rightarrow \infty$
are taken.



\begin{figure}[htbp]
  \begin{center}
    \includegraphics[width=8cm]{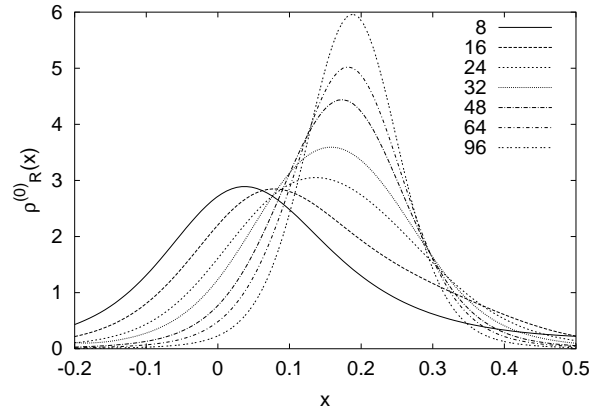}
    \caption{The distribution $\rho _{\rm R} ^{(0)} (x)$
of $\nu_{\rm R}$ in the phase-quenched model is plotted
for $\mu = 0.2$ at various $N$.
}
    \label{fig:rho0R}
  \end{center}
\end{figure}

Let us next turn to $\rho _{\rm R} ^{(0)} (x)$.
In Fig.\ \ref{fig:rho0R} we plot it for various $N$ at $\mu = 0.2$.
At small $N$ the distribution is peaked near the origin
and the dependence on $N$ is small.
As we go to larger $N$ the peak moves to $x \sim 0.2$
and starts to grow.
The VEV $\langle \nu_{\rm R} \rangle _0$,
which represents the position of the peak,
is therefore close to zero at small $N$, and it approaches
$\vev{\nu_{\rm R}}_0 = \mu$ in the large $N$ limit.

\begin{figure}[htbp]
  \begin{center}
    \includegraphics[width=8cm]{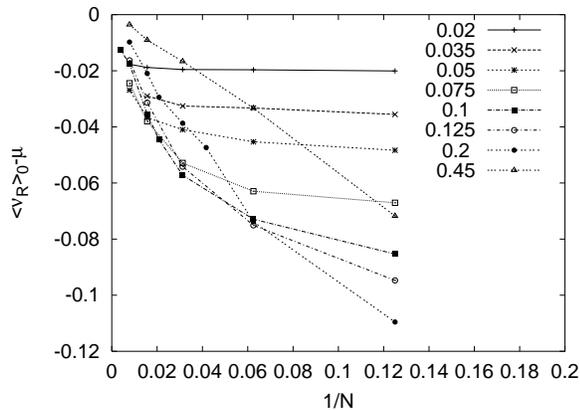}
    \caption{
We plot $\langle \nu_{\rm R} \rangle _0 - \mu$ 
against $1/N$ for various $\mu$.
}
    \label{fig:nuR0-mu}
  \end{center}
\end{figure}

In Fig.\ \ref{fig:nuR0-mu} we plot 
$\langle \nu_{\rm R} \rangle _0 - \mu$ 
against $1/N$ for various $\mu$.
We see that the large $N$ asymptotic behavior is given by
\beq
\langle \nu_{\rm R} \rangle _0  = \mu
 - C(\mu) \frac{1}{N} + \cdots  \ ,
\label{nuRlargeN}
\eeq
where the coefficient $C(\mu)$ grows as $\sim \frac{0.2}{\mu}$
for $\mu \rightarrow 0$.
From this one finds that the large $N$ scaling sets in at
%
\beq
N \gg N_{\rm tr} \simeq \frac{0.2}{\mu^2} \ .
\label{Ntransition}
\eeq



\begin{figure}[htbp]
  \begin{center}
    \includegraphics[width=8cm]{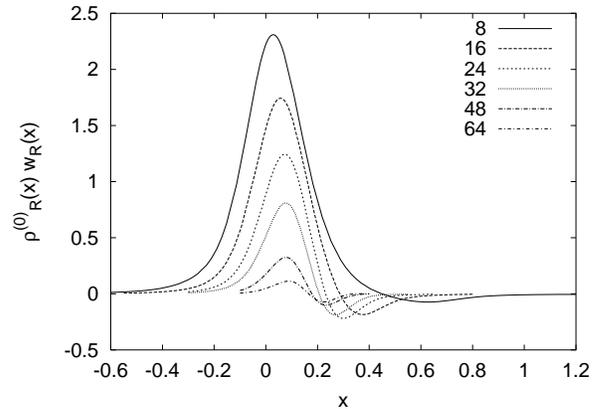}
    \caption{
The product $\rho^{(0)}_{\rm R}(x) w_{\rm R}(x)$,
which gives the unnormalized distribution 
for $\nu _{\rm R}$ in the full model,
is plotted for $\mu = 0.2$ at various $N$.
}
    \label{fig:rhoR_full}
  \end{center}
\end{figure}

The product $\rho^{(0)}_{\rm R}(x) \, w_{\rm R}(x)$
gives the unnormalized distribution of $\nu _{\rm R}$
in the full model,
which we plot in Fig.\ \ref{fig:rhoR_full}.
The distribution itself, even after appropriate normalization, 
has the non-commutativity which is inherited from
$\rho^{(0)}_{\rm R}(x)$ and $w_{\rm R}(x)$.
However, 
the VEV $\langle \nu_{\rm R} \rangle$, which is the first moment
of the distribution,
is always close to zero as one can see from Table \ref{t:1}.
The reason depends on whether $\mu \ll \frac{1}{\sqrt{N}}$ or
$\mu \gg \frac{1}{\sqrt{N}}$.
In the former case the distribution is peaked around the origin, 
which makes the first moment close to zero.
In the latter case the positive and negative regions of 
the distribution
cancel each other in the calculation of the first moment.
Thus the non-commutativity cancels in the end result
for the VEV $\langle \nu_{\rm R} \rangle$.
The situation for the imaginary part $\langle \nu_{\rm I} \rangle$
is discussed in the Appendix.

\begin{table}[ht]
\begin{center}
\begin{tabular}{| c | c | c | c |  }
\hline\hline
$N$ &$\vev{\nu_{\rm R}}$&$i \, \vev{\nu_{\rm I}}$
&$\vev{\nu}$\\
\hline
8  & 0.0056(6) & -0.1970(5)  & -0.1915(7) \\
16 & 0.0060(4) & -0.1905(13) & -0.1845(13)\\
24 & 0.0076(9) & -0.1972(14) & -0.1896(17)\\
32 & 0.0021(8) & -0.1947(19) & -0.1927(25)\\
48 & 0.0086(37)& -0.2086(54) & -0.2000(88)\\
\hline\hline
\end{tabular}
\end{center}
\caption{The VEVs 
$\vev{\nu_{\rm R}}$, $i \, \vev{\nu_{\rm I}}$ and 
$\vev{\nu}$
obtained by the factorization method for $\mu = 0.2$ at various $N$.
Statistical errors computed by the jackknife method are shown.
}
\label{t:1}
\end{table}

\section{Discussions and Future Prospects}
\label{sec:concl}

The fact that 
$\rho _{\rm R} ^{(0)} (x)$ and $\vev{\nu_{\rm R}}_0$
have the non-commutativity for
the limits $\mu \rightarrow 0$ and $N \rightarrow\infty$
can be understood as a property of 
the phase-quenched partition function (\ref{part_absdef}). This model
undergoes a phase transition to a phase of condensed Goldstone bosons
with nonzero baryon number at $\mu = m_\pi/2$ 
\cite{Gocksch:1987ha,Stephanov:1996ki}. For zero quark mass
this phase transition takes place at $\mu = 0$. If we take the thermodynamic
limit before the $\mu\to 0$ limit, the observables are calculated
in a Bose-condensed phase. If the limits are taken in the opposite 
order, the ground state is the regular chirally broken vacuum state
without Bose condensate. For this reason we obtain different
values for observables depending on the order of the limits. 
For example, if the chiral condensate $\langle \Sigma \rangle_0$ 
is calculated before the  $\mu \to 0$ limit, 
we find $\langle \Sigma \rangle_0 = 0$, whereas if we take the
$\mu \to 0 $ limit first, we find
$\langle \Sigma \rangle_0 \neq 0$. Another example is $\frac{\del}
{\del \mu} \langle \nu \rangle _0$, 
which vanishes if the $\mu \to 0$ limit is taken
before the thermodynamic limit, while it is equal to unity if the limits
are taken in the opposite order.

These two domains are separated
by the weak non-hermiticity limit \cite{Fyodorov:1996sx}, 
where the thermodynamic limit is taken at fixed $\mu^2 N$. This is 
the domain where the real part of the eigenvalues is of the order
of the average spacing of the eigenvalues (the eigenvalues are purely
imaginary for $\mu = 0$), and quantities such as the average spectral
density of the Dirac operator can be obtained analytically in this 
limit \cite{Splittorff:2003cu}.

Mathematically, the boundary $\mu^2 N$ gives
the ``average'' radius of convergence of the perturbative expansion of
the fermion determinant in powers of $ \mu$. According to Kato's
criterion the perturbative series is convergent if the norm of
the perturbative operator is less than the level spacing. Indeed
the norm of $\mu \gamma_4$ is $\mu^2$ and the average level spacing
of random matrix Dirac operator (\ref{defD}) is $\sim 1/N$. 

The Bose condensed phase is not present in the full partition function
(\ref{rmtdef}). Observables depend smoothly on $\mu$ for 
$\mu < \mu_c \ne 0$. 
Apparently the phase oscillations of the fermion determinant
completely wipe out this phase transition. 
However, because reweighting methods are 
based on the phase-quenched partition
function, we see traces of the non-commutativity of the 
$\mu \to 0$ limit and the thermodynamic limit at intermediate steps of
the calculation.

It is expected that such non-commutativity
appears also when one studies real QCD at finite baryon density
by reweighting type methods,
and the transition occurs
when the system size becomes larger than 
$V_{\rm tr} \simeq \frac{\rm const.}{\mu^2}$.
The system size 
in the recent works at small $\mu$ 
\cite{Fodor:2001au,Fodor:2001pe,Allton:2002zi,%
deForcrand:2002ci,D'Elia:2002gd} 
may be below the transition point,
but the transition will occur if one goes to larger $\mu$
for the same system size.

Since the non-commutativity cancels in the end results
for physical observables in the full model, 
the transition is not a physical one,
but it should rather be considered 
as a property of the reweighting type methods.
We hope that our results
will be useful when one tries to go beyond the small $\mu$ regime
in real QCD.
It should also be mentioned that 
the conjectured superconducting phase may be easier to
access from the other extreme, namely from the large $\mu$ regime,
where the fluctuation of the phase becomes milder again
according to the results in RMT \cite{rmt}.



\begin{acknowledgments}
We are grateful to Misha Stephanov for valuable comments.
J.A.\ acknowledges the support by the
EU network on ``Discrete Random Geometry'', grant HPRN-CT-1999-00161
as well as  by {\it MaPhySto}, Network of Mathematical Physics
and Stochastics, funded by a grant from Danish National Research Foundation.
K.N.A.'s research was partially supported by RTN grants
HPRN-CT-2000-00122, HPRN-CT-2000-00131 and HPRN-CT-1999-00161 and the INTAS
contract N 99 0590.
The work of J.N.\ was supported in part by Grant-in-Aid for 
Scientific Research (No.\ 14740163) from 
the Ministry of Education, Culture, Sports, Science and Technology.
J.V.\ was supported in part by U.S.\ DOE Grant No.\ DE-FG-88ER40388.

\end{acknowledgments}

\section{Appendix}

\begin{figure}[htbp]
  \begin{center}
    \includegraphics[width=8cm]{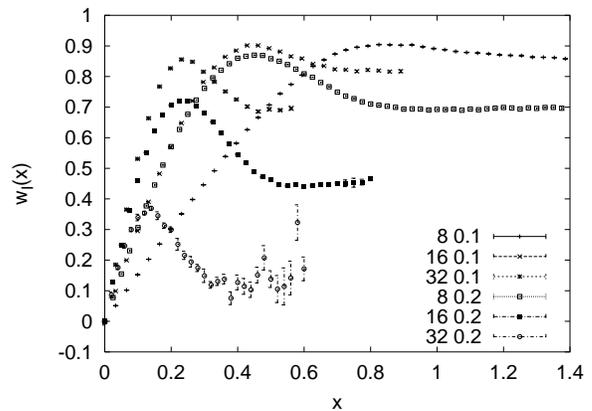}
    \caption{
The weight factor $w_{\rm I}(x)$ is plotted
for $\mu = 0.1$ and 0.2 at $N= 8,16,32$.
}
    \label{fig:wI}
  \end{center}
\end{figure}

In this Appendix we briefly describe the situation with
the imaginary part  $\nu_{\rm I}$ of the quark number density.
The weight factor $w_{\rm I}(x)$ becomes
$w_{\rm I}(x)\equiv 0 $ at $\mu = 0$ for any $N$,
and it also becomes $w_{\rm I}(x) \equiv 0$ 
in the large $N$ limit for any $\mu \neq 0$.
However the two extreme cases are not connected smoothly.
In Fig.\ \ref{fig:wI} we plot
the weight factor $w_{\rm I}(x)$
for $\mu = 0.1$ and $\mu = 0.2$ at $N=8,16,32$.

\begin{figure}[htbp]
  \begin{center}
    \includegraphics[width=8cm]{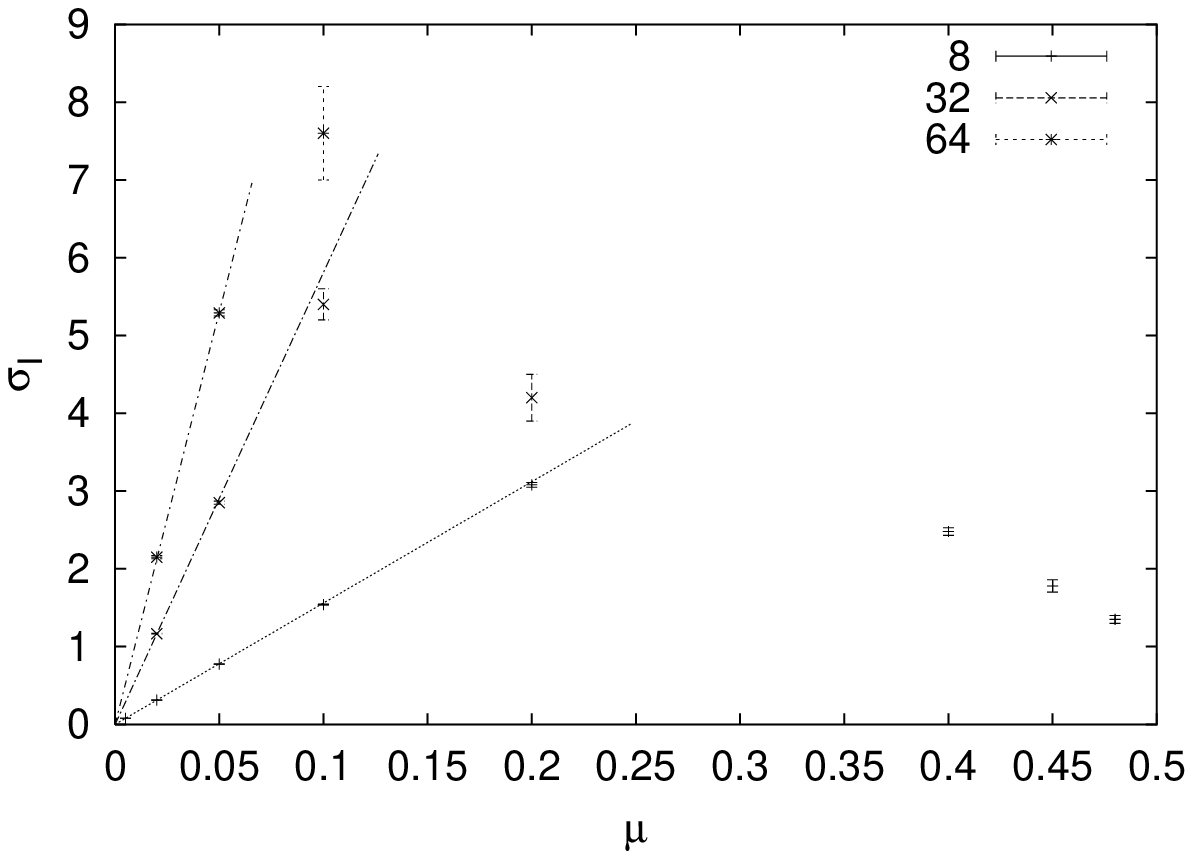}
    \caption{
The slope $\sigma_{\rm I}(\mu,N)$
of the weight factor $w_{\rm I}(x)$ in the linear regime
is plotted against $\mu$ for $N= 8,32,64$. 
For sufficiently small $\mu$,
it fits well to $\sigma_{\rm I}(\mu,N) \sim - \alpha_{\rm I} (N) \, \mu$.
}
    \label{fig:slopeI}
  \end{center}
\end{figure}

\begin{figure}[htbp]
  \begin{center}
    \includegraphics[width=8cm]{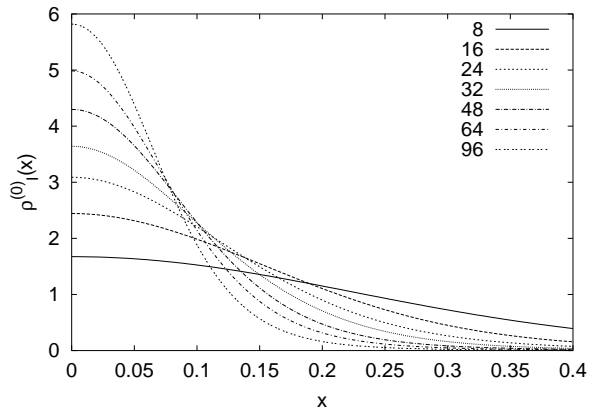}
    \caption{The function $\rho _{\rm I} ^{(0)} (x)$
is plotted 
for $N=8$, $16$, $24$, $32$, $48$, $64$, $96$ at $\mu = 0.2$.
}
    \label{fig:rho0I}
  \end{center}
\end{figure}

Since $w_{\rm I}(x)$ is an odd function due to symmetry,
it crosses the origin.
A linear regime is seen to extend from the origin as $\mu$ goes to zero
for fixed $N$.
We extract the slope in the linear regime, and plot it
as a function of $\mu$ in Fig.\ \ref{fig:slopeI}.
At small $\mu$, the slope can be fitted nicely by
\beq
\sigma_{\rm I}(\mu,N) \sim  - \alpha_{\rm I} (N)\,  \mu \ ,
\eeq
where the coefficient $\alpha_{\rm I}(N)$ grows linearly with $N$.
Thus we find that the weight factor $w_{\rm I}(x)$
has similar non-commutativity as $w_{\rm R}(x)$.

On the other hand, $\rho _{\rm I} ^{(0)} (x)$ does not 
depend much on $\mu$, and the peak at $x=0$ grows smoothly 
with $N$ as one can see from Fig.\ \ref{fig:rho0I}.
The end result for $i \, \langle \nu_{\rm I} \rangle $ does not
have the non-commutativity (See Table \ref{t:1}),
but the cancellation in this case occurs between the numerator 
and the denominator of (\ref{reweight}),
which makes it less obvious than 
the situation with $\langle \nu_{\rm R} \rangle $.


\bibliography{apssamp}

\end{document}